\documentclass[twoside,reqno]{HERON}
\usepackage{epsfig,cite,colordvi}
\usepackage{graphicx}
\usepackage{url}
\usepackage{amssymb,amsmath,amscd,epsf}
\usepackage{times}
\usepackage{makeidx}
\begin{document}

\title{Davydov-Chaban Hamiltonian within the formalism of deformation-dependent effective mass for Davidson potential}

\runningheads{Preparation of Papers for Heron Press Science Series
Books}{P. Buganu, M. Chabab, A. El Batoul, A. Lahbas, M. Oulne}

\begin{start}

\coauthor{P. Buganu}{2}, \coauthor{M. Chabab}{1}, \coauthor{A. El Batoul}{1}, \author{A. Lahbas}{1}, \coauthor{M. Oulne}{1}

\address{High Energy Physics and Astrophysics Laboratory, Faculty of Sciences Semlalia,
Cadi Ayyad University, P. O. B. 2390, Marrakesh 40000, Morocco}{1}
\address{Department of Theoretical Physics, National Institute for Physics and Nuclear Engineering,
Str. Reactorului 30, RO-077125, POB-MG6, Bucharest-Magurele, Romania}{2}

\begin{Abstract}
In this work, we modify the Davydov-Chaban Hamiltonian describing the collective motion of a $\gamma$-rigid atomic nucleus  by allowing the mass to depend on nuclear deformation. Exact analytical expressions are derived for energy spectra as well as normalized wave functions for  Davidson potential. The model, called  Z(4)-DDMD (Deformation Dependent Mass with Davidson potential), is achieved by using the Asymptotic Iteration Method (AIM).  The numerical calculations for energy spectra and $B(E2)$ transition probabilities are compared to the experimental data of  $^{192-196}$Pt isotopes. The obtained results show an overall agreement with the experiment and an important improvement in respect to other models.
\end{Abstract}
\end{start}
\section{Introduction}
The theoretical study of  excited collective states in nuclei is of particular interest  to understand shape phase transitions in nuclei. Therefore, different approaches have been developed in this context particularly in the framework of the Bohr-Mottelson Model (BMM) \cite{BMM,buga16} and of the Interacting Boson Model (IBM) \cite{IBM}. Moreover, a version of the Bohr Hamiltonian was proposed where the mass term is allowed to depend on the $\beta$ deformation variable \cite{bonat11,bonat13,chabab2016nuclear}. The newly introduced Deformation-Dependent Mass Formalism (DDMF) \cite{bonat10} offers a remedy to the problematic behaviour of the moment of inertia in the Bohr Hamiltonian, where it appears to increase proportionally to $\beta^2$ \cite{Ring}. Another direction of research was to investigate such phenomena by imposing a $\gamma$-rigidity as in the case of Z(4) \cite{Bonatsos2005} or X(3) \cite{bonatsos2006x}. These latter examples are obtained from the Davydov-Chaban Hamiltonian \cite{dav60} with an Infinite Square Well (ISW) potential for $\beta$ variable and with $\gamma$ equals to $\pi/6$ and $0$. Some improvements can be achieved by including  diverse potentials for describing $\beta$-vibrations \cite{buga16,fort},  for example, the harmonic oscillator \cite{budaca2014harmonic}, the sextic potential \cite{buganu2015analytical,budaca2016extended,Raduta:2013db}, the quartic oscillator potential \cite{budaca14}. Recently, for the first time this Hamiltonian has been used with the minimal length formalism in nuclear structure \cite{chabab2016gamma,hassan1,hassan2}.

In the present work the attention is focused on the study of the quadrupole collective states in $\gamma$-rigid case, by modifying Davydov-Chaban Hamiltonian in the framework of DDMF \cite{quesne2004,CLO15P} with Davidson  potential \cite{DAV32} for $\beta$-vibrations. The expressions for the  energy levels as well as for the wave functions  are obtained in closed analytical form by means of the Asymptotic Iteration Method (AIM)\cite{AIM03}, an efficient technique that we have used to solve many similar problems \cite{CO10,chabab2012exact,CHABAB:2012qd,chabab2015exact,chabab2016closed,clo5,clo6,celo5}. 

The Z(4)-DDMD model will be introduced in Section 2. The exact separation of the Hamiltonian and solution of angular equation are achieved in Section 3. The analytical expressions for the energy levels of Davidson potential and the wave functions are given in Section 4. Finally, Section 5 is devoted to the numerical calculations for energy spectra, $B(E2)$ transition probabilities,  while Section 6 contains our conclusions. 
\section{The Z(4)-DDM model}

In the model of Davydov and Chaban \cite{dav60}, the nucleus is assumed  to be  $\gamma$ rigid. Therefore, the Hamiltonian depends on four variables $(\beta,\theta_i)$ and has the following form \cite{dav60}
\begin{equation}
    H=-\frac{\hbar^2}{2B}\left[\frac{1}{\beta^{3}}\frac{\partial}{\partial\beta} {\beta^3}\frac{\partial}{\partial\beta}-
   \frac{1}{4\beta^2}\sum_ {k=1,2,3}\frac{Q_{k}^{2}}{\sin^2(\gamma-\frac{2}{3}\pi k)}\right]
  +V(\beta),
  \label{eq6}
\end{equation}
where $B$ is the mass parameter, $\beta$ the collective coordinate and $\gamma$ a parameter, while $Q_{k}$ are the components of the total angular momentum in the intrinsic frame and $\theta_i$ the Euler angles.

In order to construct the Davydov-Chaban equation with a mass depending on the deformation coordinate $\beta$, we follow the formalism described in Sec. II of \cite{bonat11} and to consider,
\begin{equation}
B(\beta)=\frac{B_0}{f(\beta)^2}
\label{eq7}
\end{equation}
where $B_0$ is a constant. Since the deformation function $f(\beta)$ depends only on the radial coordinate $\beta$, then only the $\beta$ part of the resulting equation will be affected.  The resulting equation reads as,

\begin{multline}
 \left[ -\frac{\sqrt{f}}{\beta^3}\frac{\partial}{\partial\beta} {\beta^3f}\frac{\partial}{\partial\beta}\sqrt{f}+
  \frac{f^2}{4\beta^2}\sum_ {k=1,2,3}\frac{Q_{k}^{2}}{\sin^2(\gamma-\frac{2}{3}\pi k)} \right]\Psi(\beta,\Omega)\\+v_{eff}\Psi(\beta,\Omega)=\epsilon\Psi(\beta,\Omega)  \label{eq8}
\end{multline}
with,
\begin{equation}
v_{eff}=v(\beta)+  \frac{1}{4}(1-\delta-\lambda)f\bigtriangledown^2f
+\frac{1}{2}\left(\frac{1}{2}-\delta\right)\left(\frac{1}{2}-\lambda \right)(\bigtriangledown f)^{2}
    \label{eq9}
\end{equation}
 where the reduced energies and  potentials are defined as $\epsilon=\frac{B_0}{\hbar^2}E$, $v(\beta)=\frac{B_0}{\hbar^2}V(\beta)$, respectively.
\section{Exactly separable form of the Davydov-Chaban Hamiltonian}

Considering a total wave function of the form  $\Psi(\beta,\Omega)=\chi(\beta)\phi(\Omega)$, where $\Omega$ denotes the rotation Euler angles ($\theta_1$,$\theta_2$,$\theta_3$), the separation of variables gives  two equations
  \begin{multline}
$\Bigg [$-\frac{1}{2}\frac{\sqrt{f}}{\beta^3}\frac{\partial}{\partial\beta} {\beta^3f}\frac{\partial}{\partial\beta}\sqrt{f}+ \frac{f^2}{2\beta^2}\Lambda +  \frac{1}{4}(1-\delta-\lambda)f\bigtriangledown^2f
$\Bigg]$\chi(\beta)\\+\frac{1}{2}$\Bigg [$$\Big($\frac{1}{2}-\delta$\Big)$$\Big($\frac{1}{2}-\lambda$\Big)$(\bigtriangledown f)^{2}+ v(\beta) $\Bigg]$\chi(\beta)=\epsilon\chi(\beta),  \label{eq10}
\end{multline}
\begin{equation}
\left[\frac{1}{4}\sum_ {k=1,2,3}\frac{Q_{k}^{2}}{\sin^2(\gamma-\frac{2}{3}\pi k)}-\Lambda\right]\phi(\Omega)=0.
    \label{eq11}
\end{equation}
where $\Lambda$ is  the eigenvalue for the equation of the angular part. In the case of $\gamma=\pi/6$, the angular momentum term can be written as \cite{Mey75},
\begin{equation}
\sum_ {k=1,2,3}\frac{Q_{k}^{2}}{\sin^2(\gamma-\frac{2}{3}\pi k)}= 4(Q^2_1+Q^2_2+Q^2_3)-3Q^2_1. \label{eq12}
\end{equation}
Eq. \eqref{eq11} has been solved by Meyer-ter-Vehn \cite{Mey75}, with the results
\begin{equation}
\Lambda=L(L+1)-\frac{3}{4}\alpha^2, \label{eq13}
\end{equation}
\begin{equation}
\phi(\Omega)=\phi^L_{\mu,\alpha}(\Omega)=\sqrt{\frac{2L+1}{16\pi^2(1+\delta_{\alpha,0})}}\left[ \mathcal{D}^{(L)}_{\mu,\alpha}(\Omega) +(-1)^L\mathcal{D}^{(L)}_{\mu,-\alpha}(\Omega) \right], \label{eq14}
\end{equation}
where $\mathcal{D}(\Omega)$ denotes Wigner functions of the Euler angles, $L$ is the total angular momentum quantum number, $\mu$ and $\alpha$   are the quantum numbers of the
projections of angular momentum on the laboratory fixed $z$-axis and the body-fixed $x'$-axis, respectively. In the literature, for the triaxial shapes, it is customary to insert the wobbling quantum number $n_w$ instead of $\alpha$, with $n_w=L-\alpha$ \cite{Mey75}. Within this convention, the eigenvalues of the angular part become :
\begin{equation}
\Lambda=L(L+1)-\frac{3}{4}(L-n_w)^2. \label{eq15}
\end{equation}

\section{Z(4)-DDM solution for $\beta$ part of the Hamiltonian}
The $\beta$-vibrational states of the triaxial nuclei, having a $\gamma$ rigidity of $\pi/6$, are determined by the solution of the radial Schr\"odinger equation
\begin{multline}
\frac{1}{2}f^2\chi'' +\left(\frac{3f^2}{2\beta}+ff'\right)\chi' +\left(\frac{3ff'}{4\beta}+\frac{(f'^2)}{8}+\frac{ff''}{4}   \right)\chi\\-\frac{f^2}{2\beta^2}\Lambda\chi+\epsilon\chi-v_{eff}\chi=0,
 \label{eq16}
\end{multline}
with
\begin{equation}
v_{eff}=v(\beta)+  \frac{1}{4}(1-\delta-\lambda)ff''
+\frac{1}{2}\left(\frac{1}{2}-\delta\right)\left(\frac{1}{2}-\lambda \right)( f')^{2}.
    \label{eq17}
\end{equation}
Setting the standard transformation of the radial wave function as $\chi(\beta)=\beta^{-3/2}R(\beta)$, we get
\begin{equation}
f^2R''+2ff'R'+(2\epsilon-2u_{eff})R=0
    \label{eq18}
\end{equation}
with
\begin{equation}
u_{eff}=v_{eff}+\frac{f^2}{2\beta^2}\Lambda+\left(\frac{3ff'}{4\beta}+\frac{3f^2}{8\beta^2}-\frac{(f')^2}{8}-\frac{ff''}{4} \right).
    \label{eq19}
\end{equation}
 Now, we consider the special case of the Davidson potential \cite{DAV32}
\begin{equation}
v(\beta)=\beta^2+\frac{\beta_0^4}{\beta^2},
    \label{eq20}
\end{equation}
where $\beta_0$ represents the position of the minimum of the potential.

According to the specific form of the potential \eqref{eq20}, we choose the deformation function in the following special form
\begin{equation}
f(\beta)=1+a\beta^2, \hspace{1.5cm}  a<<1.    \label{eq21}
\end{equation}
By inserting  the potential and the deformation function in Eq. \eqref{eq18}, one gets
\begin{equation}
2u_{eff}(\beta)=k_2\beta^2+k_0+\frac{k_{-2}}{\beta^2},   \label{eq22}
\end{equation}
with
\begin{align}
k_{2\  }=&2+a^2\Big[ (1-\delta-\lambda)+ (1-2\delta)(1-2\lambda)+\frac{7}{4}+\Lambda      \Big],   \nonumber&
\nonumber\\
k_{0\  }=& a\Big[(1-\delta-\lambda)+\frac{7}{2} +2\Lambda\Big],   \nonumber&
\nonumber\\
k_{-2}=& \Lambda+\frac{3}{4} +2\beta_0^4. &
  \label{eq23}
\end{align}
In order to apply the asymptotic iteration method  \cite{AIM03}, we propose an appropriate physical wave function in the forme :
\begin{equation}
R_{n_{\beta}L}(y)=y^{\rho}(1+ay)^{\nu}F_{n_{\beta}L}(y), \hspace{1.5cm} y=\beta^2,   \label{eq24}
\end{equation}
with
\begin{align}
\rho=&\frac{1}{4}(1+\sqrt{1+4k_{-2}})  \nonumber,&
\nonumber\\
\nu=& -\frac{1}{2}\sqrt{k_{-2}+\frac{2\epsilon}{a}-\frac{k_0}{a}+\frac{k_2}{a^2}}. &
  \label{eq25}
\end{align}
For this form,  the radial wave equation reads
\begin{multline}
F''(y)=-\Bigg[\frac{2+\rho+a(4+2\nu+\rho)y}{2y(1+ay)}\Bigg]F'(y)\\-\Bigg[ \frac{a(2\nu+\rho+3)(2\nu+\rho+1)-\frac{4k_2}{a}}{16y(1+ay)}\Bigg]F(y).
\label{eq26}
\end{multline}
We get the generalized formula of the radial energy spectrum,
\begin{equation}
\epsilon_{n_{\beta}n_{w}L}=\frac{1}{2}\left[k_0+\frac{a}{2}(3+2p+2q+pq)+2a(2+p+q)n_{\beta}+4an_{\beta}^2 \right],
\label{eq27}
\end{equation}
where $n_{\beta}$ is the principal quantum number of $\beta$ vibrations, with :
\begin{equation}
p=\sqrt{1+4k_{-2}}, \hspace{1.5cm} q=\sqrt{1+4\frac{k_2}{a^2}}.
\label{eq28}
\end{equation}
The excited-state wave functions  read,
\begin{equation}
F(y)=N_{n_{\beta}L}\ _{2}F_1\left[-n_{\beta},n_{\beta}-\frac{q}{2};-2n_{\beta}-\frac{(q+p)}{2};1+ay      \right],
\label{eq40}
\end{equation}
with 
\begin{equation}
N_{n_{\beta}L}=\left(a^{p/2+1}n_{\beta}! \ q \right)^{\frac{1}{2}} \left[  \frac{\Gamma(n_{\beta}+\frac{q+p}{2}+1)}{\Gamma(n_{\beta}+\frac{q}{2}+1)\Gamma(n_{\beta}+\frac{p}{2}+1)} \right]^{\frac{1}{2}}.
\label{eq42}
\end{equation}
The quantities $k_2$, $k_0$, $k_{-2}$ are given by Eq. \eqref{eq23}, while $\Lambda$ is the eigenvalue of angular part given by Eq. \eqref{eq15}. The excitation energies depend on three quantum numbers : $n_{\beta}$, $n_{w}$ and $L$, and four parameters : $a$ the deformation mass parameter, $\beta_0$ the minimum of the potential,  the free parameters $\delta$ and $\lambda$ coming from the construction procedure of the kinetic energy term \cite{roos83}. In the last part of the paper, a comparison to the experiment will be carried out by fitting the theoretical spectra to the experimental data. Finally, it will be shown that the predicted energy levels turn out to be independent of the choice made for $\delta$ and $\lambda$.

\section{Numerical results :}
The Z(4)-DDM model presented in the previous sections has been applied for calculating the energies of the collective states and the reduced E2 transition probabilities for $^{192,194,196}$Pt isotopes. All bands ($i.e.$ ground state, $\beta$ and $\gamma$) are characterized by the following quantum numbers :
\begin{itemize}
 \item For gsb : $n_{\beta}=0$ and $n_w = 0$;
\item	For $\beta$ band : $n_{\beta}=1$ and $n_w = 0$;
\item	For $\gamma$ band : $n_{\beta}=0$  and $n_w = 2$ for even $L$ levels and $n_{\beta}=0$  and $n_w = 1$ for odd $L$ levels.
\end{itemize}
In this work the theoretical predictions for the levels, Eq. \eqref{eq27}, depending on two parameters, namely : the potential minimum $\beta_0$ and the deformation dependent mass parameter $a$. These parameters are adjusted to reproduce the experimental data by applying a least-squares fitting procedure for each considered isotope. We evaluate the root mean square (rms) deviation between the theoretical values and
the experimental data via the formula : 
\begin{equation}\label{eq54}
\sigma=\sqrt{\frac{\sum_{i=1}^N(E_i(Exp)-E_i(th))^2}{(N-1)E(2_g^+)}}.
\end{equation}
\begin{table}[b]
\caption{The energy spectra comprising the ground, $\gamma$ and $\beta$ bands obtained with our models Z(4)-DDM Davidson (D) are compared with the values taken from \cite{buganu2015analytical} and \cite{budaca2016extended} with the available experimental data \cite{A192,A194,A196}.\label{table:table1}}

{
\setlength{\tabcolsep}{3.5pt}

\begin{tabular}{ccccccccccccccccccccc}
\hline\noalign{\smallskip}
 &\multicolumn {4}{c}{ $^{192}$Pt} & &\multicolumn {4}{c}{ $^{194}$Pt}&&\multicolumn {4}{c}{ $^{196}$Pt}\\
\cline {2 -5}\cline {7 -10}\cline {12 -15}
 &Exp&D&Ref.\cite{buganu2015analytical}&Ref.\cite{budaca2016extended}&&Exp&D&Ref.\cite{buganu2015analytical}&Ref.\cite{budaca2016extended}&&Exp&D&Ref.\cite{buganu2015analytical}&Ref.\cite{budaca2016extended}\\
\noalign{\smallskip}\hline\noalign{\smallskip}

R$_{0,0,4}$ &2.479& 2.374 &2.439 & 2.396&$$ & 2.470&2.445&2.415&2.406&$$&2.465&2.362&2.513&2.481\\
 R$_{0,0,6}$&4.314 & 3.960 &3.787 & 3.834&$$ & 4.298&4.202&3.835&3.902&$$&4.290&3.968&3.709&3.701\\
 R$_{0,0,8}$&6.377 & 5.674 &5.773 & 5.761&$$ & 6.392&6.201&5.880&5.896&$$&6.333&5.770&5.579&5.559\\
 R$_{0,0,10}$&8.624 & 7.473&7.350 & 7.484&$$ &8.672&8.408&7.573&7.713&$$&8.558&7.752&6.914&6.932\\
\noalign{\bigskip}
 R$_{1,0,0}$ &3.776& 3.714 &3.397 & 3.537&$$ & 3.858&3.666&3.706&3.809&$$ &3.192&2.970&2.954&2.977\\
 R$_{1,0,2}$&4.547& 4.726 &4.995 & 5.162&$$ &4.603&4.730&5.409&5.493&$$ &3.828&4.047&4.308&4.364\\
 R$_{1,0,4}$& & 6.118  &7.002 & 7.113&$$ & $$&7.511&6.265&7.490&$$ &&5.511&6.238&6.280\\
\noalign{\bigskip}
 R$_{0,2,2}$&1.935 & 1.857 &1.653 & 1.664&$$ & 1.894&1.892&1.661&1.676&$$ &1.936&1.848&1.646&1.643\\
 R$_{0,1,3}$&2.910 & 2.620 &2.302 & 2.345&$$ & 2.809&2.711&2.332&2.378&$$ &2.852&2.608&2.249&2.252\\
 R$_{0,2,4}$&3.795 & 4.366 &4.229 & 4.200&$$ & 3.743&4.667&4.268&4.273&$$ &3.636&4.388&4.179&4.150\\
 R$_{0,1,5}$&4.682 & 4.563 &4.342 & 4.360&$$ & 4.563&4.894&4.402&4.446&$$ &4.526&4.593&4.243&4.227\\
 R$_{0,2,6}$&5.905 & 6.686 &6.358 & 6.466&$$ & $$&7.430&6.524&6.645&$$ &5.644&6.874&6.041&6.049\\
 R$_{0,1,7}$&6.677 & 6.523 &6.065& 6.215&$$ & $$&7.230&6.235&6.392&$$ &&6.694&5.737&5.754\\
 R$_{0,2,8}$&8.186 & 8.925 &9.163 & 9.203&$$ & $$&10.269&-&9.508&$$ &7.730&9.424&8.564&8.573\\
\noalign{\smallskip}\hline\noalign{\smallskip}
$\sigma$& & 0.526  & 0.614 &0.593 &$$ &&0.338&0.543&0.515&&&0.550&0.682&0.683\\
a& & 0.002  &  & &$$ &&0.058&&&&&0.122&&\\
$\beta_0$& & 1.32  &  & &$$ &&1.31&&&&&1.14&&\\

\noalign{\smallskip}\hline

\end{tabular}
}
\end{table}

This quantity represents the rms deviations of the theoretical calculations from the experiment, where $N$ denotes the number of states, while $E_i(exp)$ and $E_i(th)$ represent the theoretical and experimental energies of the i-th level, respectively. $E(2_g^+)$ is the energy of the first excited level of the ground state  band (gsb). 

From Table \eqref{table:table1}, one can see that the obtained results for the levels belonging to gsb, $\beta$ and $\gamma$ band are in quite satisfactory agreement with experimental data. Analyzing the mean deviation  corresponding for each nucleus, we can see that the present results are fairly better that those obtained by Z(4)-sextic model. This is explained by the fact that here the mass parameter depends on the $\beta$ variable, while in Refs. \cite{buganu2015analytical,budaca2016extended} the mass is considered as a constant.

Similarly, we have calculated the intraband and interband $B(E2)$ transition rates, normalized to the transition from the first excited level of the ground state band (gsb) to the ground state, using the same optimal values  of the three parameters obtained from fitting the energy ratios. From the obtained theoretical results shown in Table \eqref{table:table2}, one can remark some discrepancies within the ground state band of the higher $L$ levels, while the experimental values show a decreasing trend.  For the intra-band transition from the $\gamma$ band to the gsb,  both models give good results, while for transition from the $\beta$ band to the gsb, the agreement is only partially good.

\begin{table}[b]
\caption{The comparison of experimental data \cite{A192,A194,A196} (upper line) for several $B(E2)$ ratios of  nuclei to predictions by the Davydov-Chaban Hamiltonian with $\beta$-dependent mass for the Davidson potential (lower line), using the parameter values shown in Table \eqref{table:table1}. \label{table:table2}}

{
\setlength{\tabcolsep}{5pt}

\begin{tabular}{lllllllllll}
\hline

 nuleus & $\frac{4_g\rightarrow 2_g}{2_g\rightarrow 0_g}$ & $\frac{6_g\rightarrow 4_g}{2_g\rightarrow 0_g}$ &$\frac{8_g\rightarrow 6_g}{2_g\rightarrow 0_g}$ & $\frac{10_g\rightarrow 8_g}{2_g\rightarrow 0_g}$ & $\frac{2_{\gamma}\rightarrow 2_g}{2_{1}\rightarrow 0_g}$&$\frac{2_{\gamma}\rightarrow 0_g}{2_g\rightarrow 0_g}$&$\frac{0_{\beta}\rightarrow 2_g}{2_g\rightarrow 0_g}$&$\frac{2_{\beta}\rightarrow 0_g}{2_g\rightarrow 0_g}$\\
 &&&&&&$\times 10^3$&&$\times 10^3$ & \\
\hline

    $   ^{192}$Pt & $1.56(12)$ & $1.23(55)$&$ $ & $$&$1.91(16)$ & $9.5(9)$&$$&$$&\\
  & $1.60$ & $2.34$&$3.02$ & $3.69$&$1.63$ & $0.0$&$0.80$&$15.12$\\\\

    $   ^{194}$Pt & $1.73(13)$ & $1.36(45)$&$1.02(30) $ & $0.69$&$1.81(25)$ & $5.9(9)$&$0.01$&$$&\\
  & $1.60$ & $2.34$&$3.01 $ & $3.65$&$1.62$ & $0.0$&$1.15$&$52.33$\\\\

     $   ^{196}$Pt & $1.48(3)$ & $1.80(23)$&$1.92(23) $ & $$&$$ & $0.4$&$0.07(4)$&$0.06(6)$&\\
  & $1.68$ & $2.54$&$3.33$ & $4.10$&$1.70$ & $0.0$&$1.46$&$45.70$\\
\hline
\end{tabular}
}
\end{table}

\section{Conclusion :}
A new solution for the Davydov-Chaban Hamiltonian within the DDM for Davidson potential is proposed, called Z(4)-DDM Davidson.
From the mathematical point of view: this work is achieved through the use of  Asymptotic Iteration Method AIM to derive exact analytical expressions for the spectra and wave functions. From the physics point of view : the numerical realization of this model consisted of calculating energy spectra and transition probabilities of $^{192,194,196}$Pt isotopes using Davidson,  as collective potential, that we compare to experimental data and confront with other models calculations. Experimental intraband and interband transitions rates are slightly underestimated. 

\section*{Acknowledgements}
A. Lahbas acknowledges financial support by the High Energy Physics and Astrophysics Laboratory of Cadi Ayyad University. He would like to thank the organizing committee for the hospitality and the wonderful scientific meeting.

\end{document}